\documentclass[conference]{IEEEtran}
\usepackage[T1]{fontenc}% optional T1 font encoding
\usepackage{amsmath}
\usepackage[cmintegrals]{newtxmath}
\usepackage{url}
\usepackage{soul,color,graphicx,float,epstopdf}
\usepackage{tablefootnote,textcomp,gensymb}
\usepackage{eurosym,booktabs,multirow}
\usepackage{subcaption}
\usepackage{float}
\usepackage{amsfonts} 
\usepackage{algorithmic}
\usepackage{array}
\usepackage{stfloats}
\usepackage{mathtools}
\usepackage{pdfpages}
\usepackage{comment}
\usepackage{balance}
\usepackage{xcolor}
\usepackage[normalem]{ulem}
\usepackage{enumitem}
\usepackage[implicit=false]{hyperref}

\usepackage{tikz}
\usetikzlibrary{arrows,shapes,positioning,shadows,trees}
\tikzset{
  basic/.style  = {draw, text width=4cm, font=\sffamily, rectangle},
  root/.style   = {basic, rounded corners=2pt, thin, align=center,
                   fill=gray!20},
  level 2/.style = {basic, rounded corners=6pt, thin,align=center, fill=blue!10,
                   text width=9em},
  level 3/.style = {basic, thin, align=left, fill=yellow!5, text width=10em}
}

\definecolor{Orange}{rgb}{1,0.5,0}
\newcommand{\todo}[1]{\textsf{\textbf{\textcolor{Orange}{[#1]}}}}

\begin{document}
\title{Open RAN meets Semantic Communications: \\ A Synergistic Match for Open, Intelligent, \\ and Knowledge-Driven 6G}

\author{\IEEEauthorblockN{
Peizheng Li,
Adnan Aijaz
}\\ 
\vspace{-3.00mm}
\IEEEauthorblockA{
\IEEEauthorrefmark{0} Bristol Research and Innovation Laboratory, Toshiba Europe Ltd., U.K.\\
Email: {\{firstname.lastname\}@toshiba-bril.com}
}}
\maketitle
\begin{abstract}
\textcolor{black}{Improving sustainability, enhancing spectral and energy efficiency, and bringing in-network intelligence and reasoning are the driving forces for 6G. In this context, semantic communications (SemCom) and open radio access networks (Open RAN) are emerging as focal points of research. SemCom is widely viewed as a disruptive paradigm that creates the possibility of knowledge-driven networks. On the other hand, Open RAN paves the way for open and programmable networks with native intelligence. This paper investigates the synergies between SemCom and Open RAN and introduces the concept of semantic-aware Open RAN. It presents the main architectural components and functionalities along with some of the key applications of semantic-aware Open RAN. It also conducts demonstration and evaluation of a remote WiFi localization built on semantic-aware Open RAN. Finally, it highlights some open challenges for semantic-aware Open RAN.}

%The improvement of efficiency and intelligence are the driving forces of 6G exploration. In this context, Semantic Communications (SemCom) and open radio access networks (Open RAN) have emerged as focal points of research and development. SemCom exhibits a remarkable ability to extract meaningful semantic information from raw data through the utilisation of prior knowledge. Meanwhile, Open RAN stands out with its distinctive attributes of open, intelligent, and disaggregated architecture. This paper introduces a novel concept named semantic-aware Open RAN, aimed at seamlessly integrating the advantages of SemCom and Open RAN. We present the details of this innovative approach. The major architectural components and their functionalities, including the Semantic engine, Semantic RIC, and CU-SP modules, are elaborated on. Additionally, we highlight various potential applications that can harvest significant benefits from the adoption of semantic-aware Open RAN. To illustrate the practical workflow of the semantic-based application within the Open RAN, we demonstrated a specific sensor fusion application for localisation. Lastly, we identify the open challenges for the semantic-aware Open RAN study.
\end{abstract}


\section{Introduction}
\label{sec:introduction}
\vspace{-1.50mm}
While 5G technology is being commercialized, research activities toward 6G have started. 
%In the wake of maturing 5G networks, developing 6G communication systems has become a prominent research area. 
The radio access network (RAN) takes center stage in the 6G design. Driven by Shannon's seminal work on the information theory of communications~\cite{shannon1948mathematical}, RAN components have advanced significantly. State-of-the-art (SOTA) systems approach the Shannon limits. To further improve performance and efficiency of the RAN in 6G, to beyond Shannon level, artificial intelligence (AI) and machine learning (ML) techniques are gaining traction. Such techniques provide a data-driven approach that is capable of extracting hidden patterns in the data features. Hence, the key issues toward 6G RAN design funnel to: (a) how to process enormous data in the RAN and obtain meaningful knowledge, and (b) how to retain and leverage this knowledge to improve the RAN's operational efficiency. Semantic Communications (SemCom)~\cite{gunduz2022beyond} provides a promising approach to address them.

SemCom departs from conventional bitwise communications and aims to revolutionize the paradigm of information exchange by emphasizing semantic aspects. 
% It involves extracting semantic information from raw data for potential transmission and processing, presenting intriguing possibilities for more efficient machine-to-machine and human-machine communications. 
% The foundation of SemCom relies on the accuracy of semantic representation. Such capability is learned from the pre-collected knowledge bases (KB). 
In general, SemCom can be regarded as a procedure of leveraging the pervasive knowledge in the network. Interestingly, it implicitly follows the concept of the knowledge plane in Internet network architecture, initially proposed by Clark \emph{et al.}~\cite{clark2003knowledge}. 
% which emerged as a response to the need for a comprehensive framework of dynamically adapting and configuring networks to meet evolving service conditions, by utilising the knowledge in networks and understanding high-level instructions. This framework aimed to achieve "close the loop" control while incorporating learning and reasoning capabilities.
However, this proposal encountered difficulties in effectively representing, utilizing, and routing knowledge. Despite the rapid development of network architecture over the subsequent two decades, the issue of effectively harnessing existing knowledge remained unaddressed. While at present, SemCom emerges as a valuable tool for constructing the knowledge plane.
% from the abundant data available on users, control, and management.
So, where and how the SemCom paradigm should be integrated into the SOTA networks? Edge devices are occasionally brought up as SemCom venues for task-oriented studies~\cite{xu2023edge}. For instance, Shi \emph{et. al}. introduced a federated edge intelligence for resource-efficient semantic-aware networking~\cite{shi2021semantic}; the authors in~\cite{shao2021learning} studied task-oriented communication for edge inference, and further, investigated the multi-device cooperative inference in~\cite{shao2022task}; the authors proposed a distributed SemCom system for IoT in~\cite{xie2020lite}. Several prior studies have also examined the semantic applications in Open RAN focusing on slicing and scalability~\cite{puligheddu2022sem}. Although authors of~\cite{chaccour2022less}, and~\cite{yang2022semantic} touch on the semantic reasoning plane and the semantic-empowered network definition; however, there lacks a comprehensive discussion on how Open RAN ought to support SemCom at the protocol level. By identifying the complementary advantages of SemCom and Open RAN, 
this paper proposes the construction of semantic-aware Open RAN. 

\begin{figure*}
\centering
%\centerline{
\includegraphics[width=0.9\linewidth]{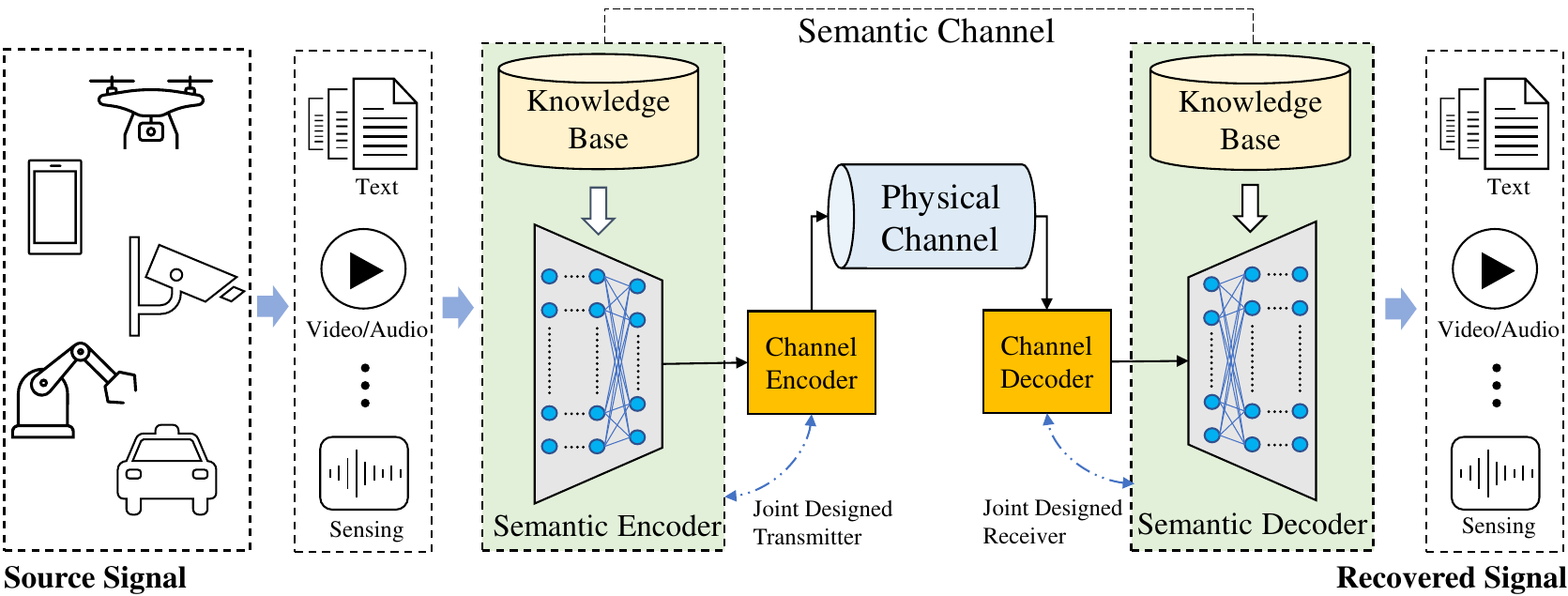}
% \vspace{-1.50mm}
\caption{Illustration of the SemCom concept.}
\label{fig: semantic communication illustration}
\vspace{-1.25mm}
\end{figure*}
\vspace{-2.25mm}

\subsection{Benefits of semantic-aware Open RAN}
% \vspace{-1.50mm}
Open RAN has been developed with the objective of fostering intelligence, openness, virtualization, and complete interoperability within networks. 
% Open RAN encompasses notable characteristics, including the incorporation of artificial intelligence (AI) through the RAN Intelligent Controllers (RICs). Moreover, it adheres to standardized and well-defined hardware and software services, effectively virtualizing network functions to the greatest possible extent, that allows for seamless communication between diverse components and systems, thereby promoting interoperability and eradicating vendor lock-in for Open RAN. 
After years of dedicated efforts in standardization and promotion, Open RAN has emerged as a highly captivating network architecture for the deployment of 5G, extending its significance to Beyond-5G and 6G implementations. 
% Open RAN is able to serve a massive user base as a ubiquitous infrastructure.
Therefore, the establishment of semantic-aware Open RAN will lead to significant benefits:
\begin{itemize}[leftmargin=*]

\item \textbf{Enhancing Network Intelligence:} SemCom will revolutionize Open RAN by creating an end-to-end sematic-aware network that enables more compacted semantic information exchange by leveraging the pertained knowledge.
% , for task processing and information interaction. 
% That allows operators with enhanced network intelligence, resulting in reduced communication overhead, and improving service quality and operational efficiency. 
This enhances network intelligence, reduces communication overhead, and improves service quality and operational efficiency for operators.

\item \textbf{Boosting Performance:} 
SemCom empowers knowledge-based reasoning in Open RAN, driving performance improvements in network optimization, anomaly detection, and fault maintenance. SemCom applications also encourage individual development for innovative solutions, that match the disaggregation and open properties of Open RAN, for fostering competition and technological advancements.

\item \textbf{Future-oriented Interoperability:} SemCom provides interoperability within Open RAN and other communication technologies, enabling efficient machine-to-machine and human-machine communication, which facilitates seamless integration with RAN, edge nodes, and IoT devices under the unified SemCom framework.
\end{itemize}

\subsection{Contributions}
To the best of our knowledge, this is one of the first works that systematically discusses the integration approach of Open RAN and SemCom at all stacks, and details the structural enhancements required by Open RAN to accommodate SemCom's new features and applications. Our main contributions are summarized as follows. 
\begin{itemize}[leftmargin=*]
    \item We articulate the synergies of SemCom and Open RAN, and identify the limitations of current Open RAN in supporting SemCom. 
    \item We propose a holistic SemCom-aware Open RAN architecture.
    \item We list potential applications benefiting from the semantic-aware Open RAN, and a particular remote localization application is discussed to demonstrate the semantic workflow within Open RAN.
    \item We discuss various challenges for the development of semantic-aware Open RAN.
\end{itemize}

\subsection{Paper outline}
% \vspace{-1.50mm}
The remainder of this paper is structured as follows: Sec.~\ref{sec:backbround} provides the foundational knowledge of the SemCom and Open RAN. Sec.~\ref{sec: mutual benefits} analyses their mutual benefits, and the innovative architecture is elaborated in Sec.~\ref{sec: Semantic-aware Open RAN}. The potential applications that gain from the proposed semantic-aware Open RAN are summarized in Section~\ref{sec:potential applications}.
In Sec.~\ref{sec:usecases}, a remote localization application is detailed to show the semantic workflow, and we sum up the major challenges of semantic-aware Open RAN in Sec.~\ref{sec:discussions}. Finally, the paper is concluded in Sec.~\ref{sec:conclusion}.

\section{Preliminaries}
\label{sec:backbround}
% \vspace{-1.50mm}
\subsection{SemCom}
\label{subsec:semcom}
\subsubsection{Objective of SemCom}
SemCom can be defined as a communication paradigm that prioritizes the successful transmission and delivery of semantic information contained in a source, as opposed to the transmission of raw source data bit-by-bit in traditional communication networks. The SemCom study still is in its infant stage, research efforts are given commonly in constructing substrates for semantic extraction, semantic information transmission, and semantic recovery and measurement in a robust and general manner, with the goal of improving the efficiency, reliability, and robustness of communication systems, especially in complex and dynamic environments where traditional communication methods may be limited or ineffective. 
% SemCom works through the way that detects and extracts the semantic content of the source signal and compresses or removes the irrelevant information. 
The typical methods of extracting and reconstructing semantic information so far, as summarized in~\cite{yang2023secure}, include a) autoencoder, b) information bottleneck (IB), c) knowledge graph, and, d) probability graph. This paper focuses on the autoencoder in the following discussion due to its representative. 

\subsubsection{Autoencoder-based SemCom framework}
Autoencoder comprises a pair of encoder and decoder neural networks (NNs) serving semantic processing purposes. As illustrated in Fig.~\ref{fig: semantic communication illustration}, the semantic encoder is designed to transform input data, such as text, images, or audio, into a latent representation, commonly referred to as an embedding. This embedding is intended to extract the underlying semantic content of the input data and present it in a more structured and compact manner than the original input~\cite{gunduz2022beyond,shi2021semantic,niu2022paradigm}. 
In contrast, the semantic decoder is trained to utilize the output of the encoder, i.e., the embedding, to generate the desired output. This output may take various forms, including the reconstruction of the initial input data, the generation of textual descriptions based on an image's embedding, or vice versa. 
The semantic information transmission happens in between. As a transmission optimization measure, deep joint source and channel coding (JSCC) is proposed in~\cite{8723589}, which targets combining the source and channel coding into a unified NN-based coding scheme to achieve improved overall performance. As research advances, JSCC algorithms with various capabilities, including those involving feedback or variable length codes, have been proposed. A comprehensive overall regarding the SemCom technique can be found in~\cite{gunduz2022beyond}. In essence, SemCom opens up new opportunities in wireless communication. Envisioning widespread integration of SemCom into diverse communication setups, this advancement has the potential to enhance interoperability and reliability among multiple communication entities, concurrently reducing communication overhead.

\subsection{Open RAN}
Open RAN builds upon various network innovations, including intelligent control mechanisms, functional splits, network virtualization, and open interfaces. By adopting the Open RAN approach, network operators can achieve more efficient dividends, promote increased innovation, and experience improved network performance.

\begin{figure}
\centering
%\centerline{
\includegraphics[width=0.8\linewidth]{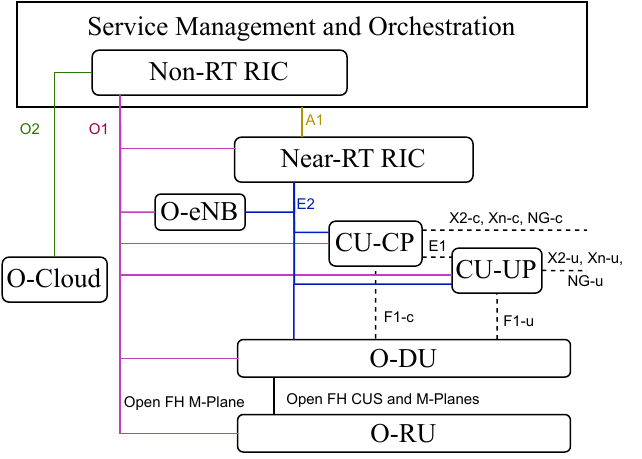}
% \vspace{-1.50mm}
\caption{Illustration of Open RAN architecture.}
\label{fig:oran architecture}
\vspace{-1.25mm}
\end{figure}

\subsubsection{Intelligence of Open RAN}
Intelligence plays a significant role in Open RAN, wherein RAN Intelligent Controllers (RICs) are crucial components. RICs enhance traditional Radio Resource Management (RRM) by incorporating data-driven approaches to process telemetries of the RAN. With the AI/ML models containerized in RICs (xApps/rApps), RICs can intelligently manage the resources in Open RAN, delivering applications such as automated network monitoring and analytic, traffic forecasting, and self-organization to enhance the quality of service.

\subsubsection{Disaggregation of Open RAN}
Open RAN inherits the 5G gNodeB splitting defined by 3GPP, consisting of the open central unit (O-CU), distributed unit (O-DU), and radio unit (O-RU). To enhance the adaptability of deploying diverse network functionalities and to minimize latency, the O-CU is further split into two logic components, the CU control plane (CU-CP) and the user plane (CU-UP). Open RAN employs the 7.2x split to disaggregate the RU and DU, and for an in-depth understanding of each constituent's functions, refer to~\cite{polese2023understanding}.

% wherein the CU-CP handles the transferring of commands from high-PHY in the DU to the low-PHY of the RU. The CU-UP handles the actual data traffic (I/Q samples) between RU and DU, Open RAN adopts the 7.2x split for the disaggregation of RU and DU, in which RU is responsible merely for FFT and cyclic prefix addition/removal operations, while DU performs all the remaining functionalities of the physical, Medium Access Control (MAC) and Radio Link Control (RLC) layers \cite{polese2023understanding}.

\subsubsection{Openness of Open RAN}
\label{subsubsec:openness of open ran}
Openness involves the adoption of open and interoperable interfaces, enabling equipment from diverse vendors to function harmoniously in Open RAN. This in turn facilitates greater flexibility for operators in network deployment and management. The hierarchical architecture of Open RAN is demonstrated in Fig.~\ref{fig:oran architecture}, where the E2 interfaces establish connections between the near-real-time(RT) RIC to the CU/DU for the near-RT control. Linking near-RT RIC and non-RT RIC, the A1 interface enables non-RT control and intelligent model updates in near-RT RIC. 
The Non-RT RIC communicates with other Open RAN components through the O1 interface, for management and orchestration. Additionally, the O2 interface connects the Non-RT RIC and the Service Management and Orchestration (SMO) to the O-Cloud. The X2-c(u), Xn-c(u), and NG-c(u) interfaces are used for information sharing for other CUs.

\section{Why Combine Open RAN and SemCom?}
\label{sec: mutual benefits}
\vspace{-1.0mm}
The synergy between SemCom and Open RAN is anticipated to yield mutually beneficial outcomes. Through exploiting the inherent properties of SemCom and Open RAN and their correlations, the potential benefits are discussed below.
% \vspace{-1.25mm}
\subsection{What SemCom Can Offer to Open RAN}
\textit{SemCom offers the possibility of evolving to knowledge-driven semantic-aware Open RAN.}
\begin{enumerate}[leftmargin=*]

\item \textbf{Transmission Compression and Resilience:} Integrating the SemCom paradigm into Open RAN can significantly reduce communication overhead due to its information compression capability. SemCom can also enhance transmission resilience by considering channel conditions during the source coding process, ensuring performance in noisy channels. Moreover, semantic-based information recovery enables reconstruction even in scenarios with partial information loss, such as hardware impediments.

\item \textbf{Improved Operational Efficiency and Interoperability} 
The reduction of communication overhead undoubtedly brings the operational efficiency improvement of Open RAN. Meanwhile, the deployment of the SemCom model at different levels within Open RAN can lead to an improvement in interoperability, as they can identify and process semantically similar information by leveraging shared semantic understanding. Moreover, its broader deployment beyond Open RAN can facilitate smooth interactions between different communication systems. 

\item \textbf{Knowledge-driven Semantic-aware Open RAN}
Beyond its data compression capabilities, SemCom can be regarded as a potent framework that effectively harnesses prior knowledge. Full stack integration of SemCom in Open RAN holds the potential to enhance the platform with knowledge-driven reasoning capabilities. This transformative paradigm shift would empower Open RAN to expedite intelligent task processing and resource allocation through knowledge distilled from historical operational records.
\end{enumerate}

\subsection{What Open RAN Can Offer to SemCom}
\textit{Open RAN offers the most suitable and extensive deployment platform for SemCom to date.}

\begin{enumerate}[leftmargin=*]
\item \textbf{Ready-made AI/ML Application Pipeline:}
Open RAN's hierarchical architecture and programmable RICs, position it as the most suitable network paradigm so far for exploring SemCom compatibility. Within the Open RAN framework, there already has been extensive research into the AI/ML development and deployment pipeline. Meanwhile, its open interfaces and accessible data flow streamline data collection for the generation of the necessary knowledge base (KB), a crucial element in training SemCom models. 

\item \textbf{Large-scale Deployment Possibility:} 
Incorporating semantic models within containers becomes feasible within the Open RAN framework, streamlining the deployment procedure. It can be expected that the larger-scale deployment of Open RAN infrastructure would boost the beyond-Shannon value of SemCom.

\item \textbf{Openness Features:} Open RAN's uniqueness lies in its openness (Sec.~\ref{subsubsec:openness of open ran}). By incorporating Open RAN, SemCom would share the openness feature of Open RAN. This may benefit the promotion and standardization of SemCom.
\end{enumerate}

\subsection{Limitations of the SOTA Open RAN}

It is emphasized that SOTA Open RAN implementations provide only partial backing for SemCom, mainly relevant to the predetermined AI/ML capability. This hinders the full realization of the semantic-aware Open RAN. 
\begin{enumerate}[leftmargin=*]
\item To enable semantic-aware Open RAN, it is important to establish coordination among the top-level timing, interfaces, and data flow pertaining to semantic tasks. Consequently, the integration of an additional SemCom engine becomes essential to fulfill this role. The designated SemCom engine would be responsible for coordinating internal data exchange and resource allocation within the Open RAN.

\item SemCom applications within Open RAN need to accommodate a diverse range of tasks. Ensuring compatibility between semantic processing and traditional task operations is crucial for the successful delivery of services.

\item Since Open RAN serves as essential infrastructure, its SemCom support ought to assume an integrating and facilitating role concerning various network domains like IoT, Edge nodes, and UEs. The aim is to ensure seamless SemCom service coverage across all these domains. This prospect has not been aware of by SOTA Open RAN architecture.
\end{enumerate}

Therefore, it can be deduced that for the future-oriented semantic-aware Open RAN, the holistic semantic control loop, the SemCom function disaggregation, semantic-empower devices, and applications interoperation are necessary.
% That asks for architectural innovations upon the SOTA Open RAN. The authors would suggest and elaborate on possible functional enhancements in the following sections.
Such advancements demand architectural innovations beyond the SOTA Open RAN.
In the subsequent sections, we will propose and elaborate on potential functional enhancements to address these requirements.

\begin{figure}
\centering
%\centerline{
\includegraphics[width=0.9\linewidth]{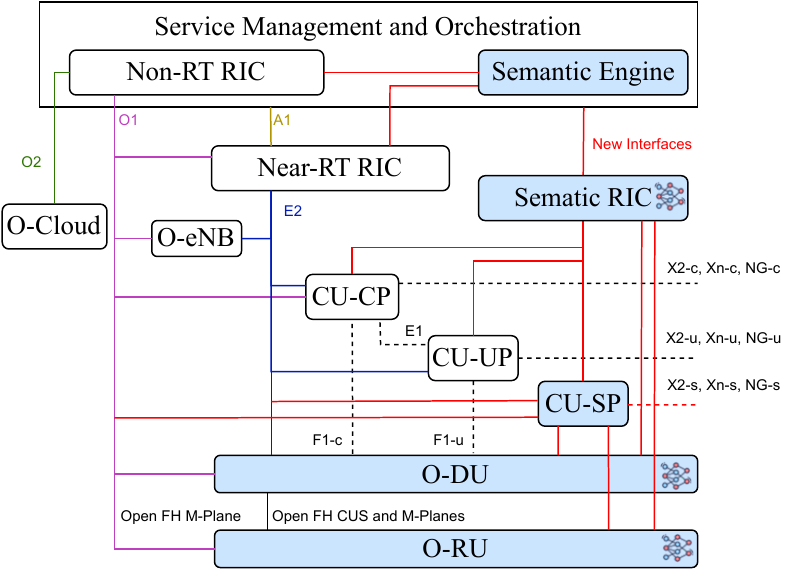}
% \vspace{-1.50mm}
\caption{The proposed Open RAN architectural enhancement for SemCom support.}
\label{fig: architecture enhancement}
\vspace{-1.25mm}
\end{figure}

\section{Semantic-aware Open RAN}
\label{sec: Semantic-aware Open RAN}
The proposed semantic-aware Open RAN architecture is holistically illustrated in Fig.~\ref{fig: architecture enhancement}, wherein three new modules, \textbf{Semantic Engine}, \textbf{Semantic RIC (S-RIC)}, and \textbf{Semantic Plane (CU-SP)}, are newly put forward for the essential support of native semantic processing capability. Meanwhile, O-DU, O-RU, and edge node/UE are suggested for extra functional enhancement. It is important to emphasize that this paper provides recommendations merely for the functional types of the new modules and involved interfaces (marked in red in Fig.~\ref{fig: architecture enhancement}). The final decision should be made by the dedicated work groups within the Open RAN alliance.

\begin{itemize}[leftmargin=*]
    \item \textbf{Semantic Engine:} The intended placement of this engine is within the network SMO framework, with the primary objective of facilitating the efficient and effective delivery of semantic services for both internal and external applications. This is achieved through the orchestration of resources for semantic information processing and management of semantic model lifecycles with user experience. The semantic engine is designed to establish connections with three distinct RICs, namely the Non-RT RIC, Near-RT RIC, and Semantic RIC. It assumes the responsibility of prioritizing and provisioning the semantic models integrated within the RICs. To maintain consistency with the existing links between the Non-RT RIC and Near-RT RIC, we propose utilizing the A1 inference for establishing the connection between the semantic engine and other RICs.
    
    \item \textbf{Semantic RIC (S-RIC):} S-RIC serves as the centralized semantic controller within the Open RAN framework, designed to be a programmable and extensible unit that facilitates the deployment of diverse custom SemCom applications. Its operations are directed by commands from the Semantic Engine, which governs its activation and potential task execution. Regarding internal connectivity, the purpose of the S-RIC is to create connections with both the CU-CP and CU-UP to oversee divided services and data. Furthermore, establishing direct connections between the S-RIC and the O-DU and also O-RU are highly desirable, as it provides a direct pathway for efficient SemCom processing, particularly beneficial for low-latency applications.
    
    \item \textbf{CU Semantic Plane (CU-SP):} The CU-SP assumes responsibility for the control and processing of semantic information. The functionalities of the CU-SP can be categorized into three aspects: (1) It is capable of receiving instructions from the semantic RIC; (2) It facilitates communication with the CU-CP and CU-UP to coordinate tasks effectively; (3) It receives, processes, and forwards the direct flow of semantic information from the enhanced O-DU and O-RU, thereby constructing a comprehensive SemCom processing loop.
    To enhance interoperability with other RAN’s CUs, it is recommended that the CU-SP be equipped with a collection of interfaces including X2-s, Xn-s, and NG-s.

    \item \textbf{Enhanced O-DU and O-RU:} In order to achieve a comprehensive semantic processing capability at all levels within Open RAN, it is imperative to enhance both the O-DU and O-RU. These enhancements should empower the O-DU and O-RU to effectively execute SemCom applications.
    To ensure seamless task coordination, these embedded applications should be jointly controlled by the logical components CU-CP, CU-UP, and CU-SP. Additionally, considering the time-sensitive nature of semantic applications, it is crucial for the O-DU and O-RU to support direct interfaces with the S-RIC.
    
    \item \textbf{Semantic-empowered UE and Edge:} The UE and edge devices that interact with the semantic-aware Open RAN are expected to possess computational and learning capabilities. These capabilities enable them to deploy a wide range of SemCom applications, primarily focusing on semantic information extraction and interpretation.
    
    \item \textbf{Knowledge database and O-cloud:} The functionality of the KB is not explicitly depicted in Fig.~\ref{fig: architecture enhancement} as its impact is pervasive across nearly every semantic processing module. The training and validation processes of semantic models rely on the support of the KB, which underscores the significance of a unified and consistent KB in enabling the successful extraction and translation of semantic information.
    However, the management of model updates and synchronization poses a substantial challenge for the operations of the semantic Open RAN. The KB should be stored and maintained by the O-cloud for universal accessibility across all RANs within the network. The specific updates pertaining to the semantic model, in consideration of the KB, must be activated and controlled by the Semantic Engine. Additionally, it is noteworthy that the semantic-empowered edge devices and UE exhibit similar requirements for model updates. 
    
    % necessitating management by the O-Cloud, potentially finished through Over-the-Air (OTA) technology.

    \item \textbf{Semantic model management:} Although not explicitly highlighted as an architectural innovation within Open RAN, effective management of semantic models is crucial for their successful deployment in the network. While the ultimate goal of SemCom is to provide a generic encoder/decoder module applicable to all tasks, this may not be immediately feasible due to limitations in valuable datasets and computational capabilities during the early stages of development. Therefore, a more practical approach involves designing separate ML models for different tasks, or task-oriented SemCom. However, this approach leads to a significant number of SemCom models that serve distinct purposes, necessitating careful consideration of lifecycle management schemes within the SemCom application pipeline. In this regard, particular attention should be given to ML operations (MLOps) to define a specific workflow for SemCom within the Open RAN framework~\cite{9931127}.
\end{itemize}

\section{Potential Applications}
\label{sec:potential applications}
The semantic-aware Open RAN is promising for various applications characterized by substantial data interchanges and demanding control requirements. %Presented below are the typical scenarios that can gain advantages from Semantic Open RAN.

\begin{itemize}[leftmargin=*]
\item  \textit{Collaborative Robotics:} Collaborative robots (Cobots) typically involve multiple robots working together or with humans towards a common goal that relies on effective communication among them. This communication encompasses a range of aspects, including task allocation and coordination, sharing sensor data, updating status, and detecting and rectifying errors. Semantic-aware Open RAN can act as a hub for efficient and timely communication enabling centralized, distributed, or hybrid control architectures for industrial or service cobots. Semantic-awareness and the native SemCom control loops enable cobots to exchange information in more efficient and easier to interpret ways, contributing to superior decision making and task performance. 
%Eventually, it facilitates seamless robot collaboration in industrial and service robotics.

\item \textit{CSI-dependent Applications:} SemCom-aware Open RAN offers significant benefits to applications with high channel state information (CSI) demands, such as massive MIMO (mMIMO), cell-free mMIMO and reconfigurable intelligent surfaces (RIS) systems. In these systems, achieving accurate CSI estimation is challenging due to the significant amount of resources required. A SemCom Open RAN can assist in CSI estimation and compression by leveraging correlations among different antenna CSI and eliminating redundant data, wherein the SemCom capability enhanced O-RU would play a constructive role. %This SemCom approach can result in reduced communication overhead and processing complexity. 
Furthermore, SemCom-aware Open RAN can support the development of joint sensing and communications (JSAC), as the CSI it processes is regarded as a critical medium for JSAC.

% \item \textit{Massive MIMO (mMIMO):} mMIMO aims to enhance the network capacity and spectrum efficiency by utilizing a large number of transmit antennas, while accurate channel estimation is critical in its performance optimization, which heavily relies on channel state information (CSI) feedback from UE to the BS. CSI estimation is quite a challenge due to the significant amount of resources required.
% The adoption of massive MIMO in Open RAN is expected in the future and the SemCom could be employed to aid in the CSI estimation and compression process by exploiting the correlations among different antennas’ CSI and removing redundant information. This SemCom approach can also be extended to cell-free mMIMO and reconfigurable intelligent surfaces (RIS) systems, to reduce communication overhead and processing complexity.

\item \textit{XR Applications:} XR refers to technologies that combine virtual reality (VR), augmented reality (AR), and mixed reality (MR). Such applications aim to create interactive and immersive experiences for users, wherein high-quality/resolution images and video are mandatory. The semantic processing applications installed in Open RAN S-RIC can be helpful in high-quality information and personalized content generation through the contextual understanding of the task. One example is using SemCom, to generate holographic content from 2D information. By leveraging advanced SemCom algorithms, Open RAN enables the analysis and interpretation of complex data sets, allowing for enhanced user experiences.

\item \textit{Remote Sensing:} Radio signal-based sensing is an active research domain. JSAC capability is also expected to be a pillar of 6G. The foundation of radio sensing lies in establishing a mapping between radio signals and corresponding physical entities. AI/ML is usually adopted for sensing purposes, and it usually requires a large amount of radio signal features. The semantic-aware Open RAN is particularly useful in such tasks, especially for remote sensing where the communication bandwidth is limited. With the semantic information reaching Open RAN through the SemCom-empowered O-RU, CU-SP, and Semantic RIC, the advanced AI/ML model can be installed in other RICs for sensing.

\item \textit{Multimodal sensor Fusion:}
Multimodal sensor fusion entails the integration of diverse observations (modalities) from a variety of multi-type sensors for a specific objective. This approach holds significant potential for enhancing sensing accuracy by harnessing a wealth of layered information.
The semantic-aware Open RAN can better serve this fusion topic. One reason is the natural information compression property of SemCom. Another reason comes from the design flexibility of the autoencoder architecture for information fusion in the latent distribution space. With more modalities of data represented in the embedding, this autoencoder architecture shows great potential in processing multi-modal semantic information~\cite{piechocki2023multimodal, pashoutani2021multi}. Hence, the UE, edge devices, and O-RU with semantic processing ability, can contribute to the fusion of diverse signal sources.

\end{itemize}

\section{Concept Demonstration}
\label{sec:usecases}
\begin{figure}
\centering
%\centerline{
\includegraphics[width=0.85\linewidth]{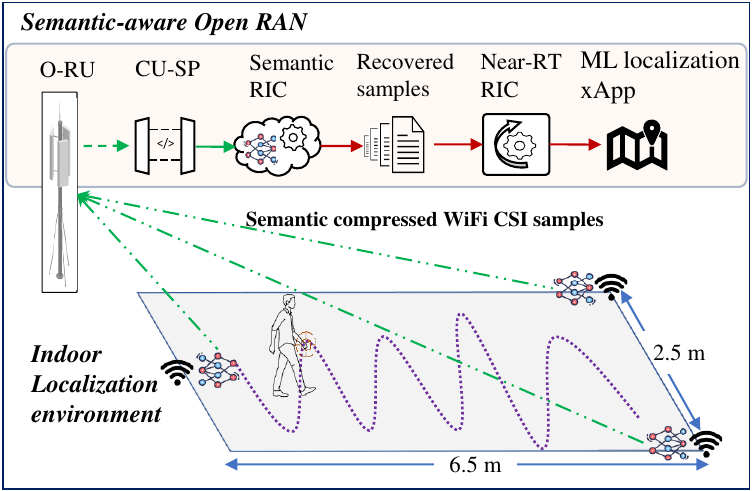}
% \vspace{-1.50mm}
\caption{Remote WiFi localisation in Semantic-aware Open RAN.}
\label{fig: wifi and mmWave fusion}
\vspace{-1.25mm}
\end{figure}
Fig.~\ref{fig: wifi and mmWave fusion} depicts how semantic-aware Open RAN can take effect in a remote localization. The process begins with edge devices compressing CSI features, and then transmitting these features to the O-RU for further processing. The CU-SP manages the flow of semantic information to the S-RIC and employs a semantic decoder to recover CSI. Subsequently, the reconstructed data is input to the Near-RT RIC, which hosts an advanced ML-based localization model for localization.

In this study, the localization feature is derived from CSI data collected from 3 WiFi access points (APs). A supervised localization NN model leveraging the raw CIS information is adopted for location estimation, whereas The data collection and preprocessing procedure are elaborated in~\cite{9412230}. The feature for each CSI sample is a combination of amplitude and phase information from 25 CSI packets of 3 APs, resulting in an input feature array of [6, 75, 30]. This signifies that the initial localization model (referred to as the raw data-based localization model) requires 13,500 features to infer a single location, which means considerable data consumption, particularly in bandwidth-restricted scenarios.

To mitigate this, the variational autoencoder (VAE)-based SemCom framework is introduced for data compression in remote localization. A conventional ResNet-18~\cite{he2016deep} and an inverted ResNet-18 serve as the encoder and decoder of the VAE. Recognizing the distribution disparities between phase and amplitude data, two separate VAE models are employed to compress each data type individually. The reconstructed amplitude and phase features are jointly fed into the Near-RT RIC boarded localization model.
\begin{figure}[t]   
    \subfloat[\label{fig: wifi CSI comrpession}]{
      \begin{minipage}[t]{0.48\linewidth}
        \centering 
        \includegraphics[width=1.6in]{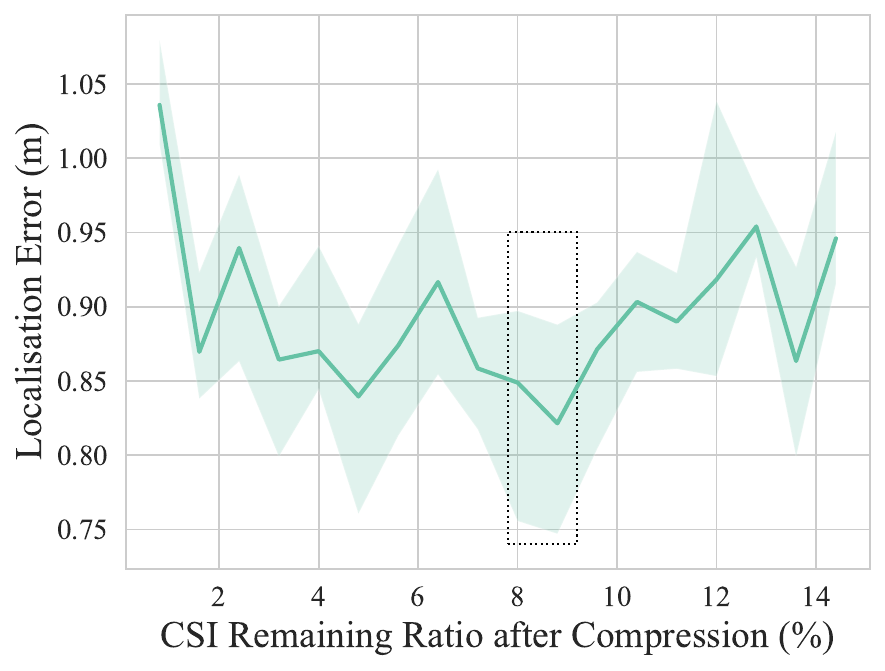}   
      \end{minipage}%
      }
      % \hfill
        \subfloat[\label{fig: CDF comparsion}]{
      \begin{minipage}[t]{0.48\linewidth}   
        \centering   
        \includegraphics[width=1.6in]{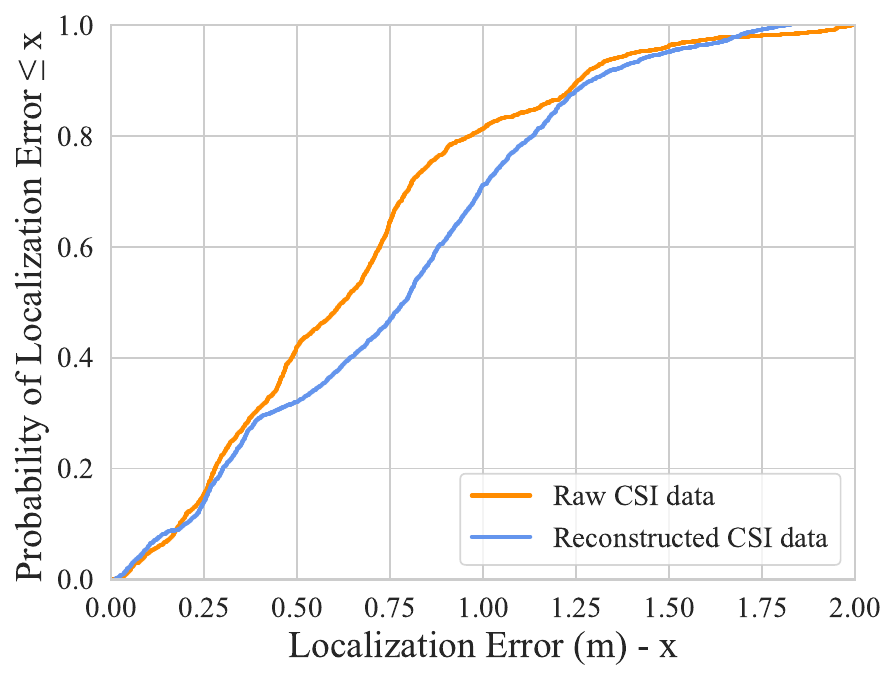}   
      \end{minipage} 
      }
      \caption{The relationship between CSI compression ratio and localization accuracy (a); The localization error CDF comparison between using the raw and reconstructed CSI data (b).
      } \label{fig:wifi results}
\end{figure}

The compression ratio and the quality of reconstruction are determined by the bottleneck size of the VAE, representing the latent space distribution. This, in turn, influences the accuracy of localization. To ascertain the optimal dimension, we vary the bottleneck size from 25 to 500, equating to a CSI remaining ratio of 0-15\% after comparison, then train the VAE and the following localization model to evaluate localization accuracy. The bounded mean plot depicting the relationship between CSI remaining ratio and localization error is shown in Fig.~\ref{fig: wifi CSI comrpession}. 
There are 12 thousand labeled samples adopted in this experiment, which are split into 80\% and 20\% for training and testing respectively. Both VAE and localization models use the Adamax optimizer.
It can be seen that the optimal performance appears at a 9\% remaining ratio of CSI, corresponding to a bottleneck size of approximately 270. Under this setting, Fig.~\ref{fig: CDF comparsion} presents the cumulative distribution functions (CDF) of localization errors, comparing the optimal SemCom-based remote localization model with the raw data-based counterpart. The average accuracy of raw data-based localization is around 0.6m, and SemCom-based remote localization is around 0.7m.
% That means, using SemCom for remote localization, optimally, the raw CSI can be compressed to one-tenth of the original size, while merely losing about 10cm of accuracy.
That means, using the VAE framework of SemCom for remote localization, under the optimal setting, the raw CSI can be compressed to one-tenth of the original size, while merely losing about 10cm of localization accuracy.

\section{Open Challenges}
\label{sec:discussions}
In this section, we highlight some of the open challenges for semantic-aware Open RAN. 
\subsection{Energy Consumption}
Decreasing the energy consumption per bit is one of the evolution trends of the modern communication system. The extra energy consumption along with SemCom consists of three parts: the SemCom model training cost, inference cost, and management cost. 
% Given the widespread deployment and utilization of semantic-aware Open RAN, 
The cumulative energy demand of widespread SemCom models can be substantial, thus presenting formidable challenges in achieving the objective of carbon neutrality. Hence, the energy-aware training and pruning strategy of NN should be explored~\cite{bullo2023sustainable}.

\subsection{Design of SemCom Models} 
The SemCom is primarily discussed in a task-oriented manner so far, resulting in the development of diverse SemCom models corresponding to different tasks. The challenge lies in achieving accurate and expeditious extraction and presentation of semantic information in the desired format. Moreover, ensuring a seamless service switch among various semantic models remains an essential concern. This potential can be satisfied by the probabilistic model selection~\cite{lotfi2022bayesian}.
%with a primary focus on achieving smooth and efficient transitions.

\subsection{Capability Evaluation of SemCom Models} 
The present analysis and testing paradigm do not provide assurances regarding the stability, reliability, security, and generalization capabilities of SemCom models, particularly when confronted with unseen or adversarial features. Given the significance of Open RAN as a critical infrastructure, it becomes imperative to devise advanced capability assessment methods for SemCom models~\cite{10107602}. These methods should address the aforementioned concerns and ensure the robust performance of the models in real-world scenarios.

\subsection{Semantic Error Correction and Model Update} 
SemCom relies on encoder and decoder pairs trained on pre-collected KB. However, input feature distribution drift can impede accurate compression by the encoder. To rectify this, a semantic error correction mechanism is required, involving semantic drift definition, monitoring, SemCom model retraining, and updates. This necessitates defining new semantic error metrics and integrating the corrections into Open RAN.

\subsection{Integration with Other AIGC models} 
The AI-generated content (AIGC) models, such as large language and image generation models, center as the most prominent research subjects nowadays. The SemCom models, to some extent, also fall within the realm of AIGC models, specifically targeting the field of communication. The integration of the SemCom model with other AIGC models can significantly enhance the capability and practicality of the SemCom-aware Open RAN, which deserves more studies.

% \subsection{Future works}

\section{Concluding Remarks}
\label{sec:conclusion}
This paper presents a pioneering approach to the semantic-aware Open RAN, designed to revolutionize the Open RAN paradigm by adopting a semantic-aware and knowledge-driven framework. This paper thoroughly elaborates on the architectural innovations added to the Open RAN, focusing on crucial components such as the semantic engine, S-RIC, and CU-SP modules. Each module's functionalities and associated interfaces are detailed. Furthermore, this paper outlines the potential applications that can leverage the benefits of the proposed semantic-aware Open RAN. To exemplify the capabilities of the proposal, a WiFi localization application is demonstrated, showcasing the workflow of semantic-aware Open RAN in action. Finally, the paper concludes the open challenges that lie ahead in the study of this domain. In the future, we will focus on addressing the above challenges and building up the testbed of SemCom-aware Open RAN, with the ultimate goal of realizing the end-to-end semantic support and holistic semantic control loop.

\section*{Acknowledgement}
\vspace{-1.00mm}
The funding support of UK government for the TUDOR project, as part of the DSIT/DCMS FONRC, is gratefully acknowledged.  %authors gratefully acknowledge the TUDOR project, part of the FRANC funded by the UK government (DSIT).

\bibliographystyle{IEEEtran} %
%\balance
%\bibliographystyle{plainnat} 
\vspace{-1.00mm}
\bibliography{IEEEabrv,references} 

% \begin{IEEEbiographynophoto}
%   {Peizheng Li} received xx
% \end{IEEEbiographynophoto}
\end{document}